\documentclass[prb,twocolumn]{revtex4-2}

\usepackage{graphicx,color,bm,amsmath,amssymb,amsfonts,amsthm}
\DeclareMathOperator{\sgn}{sgn}

\newcommand{\bfr}{{\bf r}}
\newcommand{\bfm}{{\bf m}}
\newcommand{\bfB}{{\bf B}}
\newcommand{\bfk}{{\bf k}}
\newcommand{\bfq}{{\bf q}}

\newcommand{\ua}{\uparrow}
\newcommand{\da}{\downarrow}
\newcommand{\uu}{{\uparrow \uparrow}}
\newcommand{\ud}{{\uparrow \downarrow}}
\newcommand{\du}{{\downarrow \uparrow}}
\newcommand{\dd}{{\downarrow \downarrow}}
\newcommand{\uun}{\underline{\underline n}}
\newcommand{\uuT}{\underline{\underline T}}

\newcommand{\exc}{e_{\rm xc}}

\begin{document}

\title{Spin waves with source-free time-dependent spin density functional theory}
\author{Jenna L. Bologa}
\author{Carsten A. Ullrich}
\affiliation{Department of Physics and Astronomy, University of Missouri, Columbia, Missouri 65211, USA}
\date{\today }

\begin{abstract}
Time-dependent spin density functional theory (TD-SDFT) allows the theoretical description of spin and magnetization dynamics in electronic systems from first quantum mechanical principles. TD-SDFT accounts for electronic interaction effects via exchange-correlation scalar potentials and magnetic fields, which have to be suitably approximated in practice. We consider here an approach that was recently proposed for the ground state by Sharma {\em et al.} [J. Chem. Theor. Comput. {\bf 14}, 1247 (2018)], which enforces the so-called “source-free” condition by eliminating monopole contributions contained within a given approximate exchange-correlation magnetic field. This procedure was shown to give good results for the structure of magnetic materials. We analyze the source-free construction in the linear-response regime, considering spin waves in paramagnetic
and ferromagnetic electron gases. We observe a violation of Larmor’s theorem in the paramagnetic case and a wrong long-wavelength behavior of ferromagnetic magnons.
The practical implications of these violations are discussed.
\end{abstract}

\maketitle

\section{Introduction}\label{Sec:1}

One of the most important dynamical phenomena occurring in magnetic materials are spin wave excitations, also known as magnons.
In recent years, magnons have become of great interest as potential carriers of quantum information, giving rise to the
field of magnonics \cite{Barman2021}. Another exciting recent development is the study of magnon topology \cite{McClarty2022}.

From a theoretical perspective, magnon dispersions in magnetic materials can be calculated in several ways. Widely used are
(semi)classical approaches based on model Hamiltonians \cite{Rezende2020}. Among the first-principles approaches, time-dependent spin density functional theory (TD-SDFT)
has a prominent position, with a wide and diverse range of methods and applications \cite{Savrasov1998,Capelle2001,Buczek2009,Eriksson2017,Eich2018,DAmico2019,Tancogne2020,Singh2020,Anderson2021}.

TD-SDFT can be viewed as the time-dependent generalization of spin density functional theory (SDFT) \cite{Barth1972,Gunnarsson1976,Gidopoulos2007},
which is the most commonly used {\em ab initio} approach for magnetic materials.
SDFT describes the ground state of electronic many-body systems with charge and spin degrees of freedom, which can be magnetic because of unpaired spins (such as in
open-shell molecules), due to magnetic interactions (such as in ferro- or antiferromagnets), or due to the influence of external magnetic fields that couple only to the spin.
TD-SDFT extends this to the dynamics of charge and spin fluctuations, both in the linear and nonlinear regime.

The key ingredients of (TD-)SDFT are exchange-correlation (xc) scalar potentials and magnetic fields, whose functional forms (depending on the density $n$ and magnetization
$\bf m$ as basic variables) have to be approximated in practice. Essentially all approximate xc functionals in modern DFT come in an explicitly spin-dependent format,
intended for situations in which the magnetization is collinear, i.e., the magnetization vector $\bfm(\bfr)$ points along a fixed direction in space everywhere
(conventionally taken to be $z$).
Noncollinear magnetism is usually described using a simple approximation involving local rotation of the spin-quantization axis \cite{Kubler1988,Sandratskii1998},
but other, more general SDFT approaches for noncollinear magnetism have been developed
\cite{Peralta2007,Eich2013,Scalmani2012,Bulik2013,Pittalis2017,Goings2018,Ullrich2018,Pluhar2019,Desmarais2021,Tancogne2023a,Tancogne2023b,Hill2023,Pu2023}.

In this paper, we are concerned with one particular approximation within SDFT for noncollinear systems, proposed in 2018 by Sharma {\em et al.} \cite{Sharma2018}.
This approximation restricts the xc magnetic fields of SDFT to those which do not contain any source terms, just like
any physical magnetic field, in accordance with classical Maxwell theory. For many, if not most, approximate xc functionals in SDFT this is not the case,
i.e., the resulting xc magnetic field may contain monopole source terms. To remove these contributions,
Sharma {\em et al.} proposed a construction that enforces the source-free condition for any such given approximation.
It was found that this construction improved the ground state description of a number of magnetic materials \cite{Krishna2019,Moore2024}.

Here, we extend this source-free construction into the dynamical regime, and use it within linear-response TD-SDFT to calculate spin-wave dispersions of paramagnetic and
ferromagnetic homogeneous electron gases over a wide range of parameters, and in three and two dimensions (3D and 2D). The adiabatic local spin-density approximation
(LSDA) is known to correctly describe the main physical features of spin waves: in the paramagnetic case, the long-wavelength limit is governed by
Larmor's theorem, and in the ferromagnetic case, one obtains gapless magnon dispersions with a $q^2$-behavior for small wavevectors $q$.
As we will show, the source-free construction violates both requirements. We will analyze this in detail and discuss practical implications.

This paper is organized as follows. Section \ref{Sec:2} gives the theoretical background of SDFT and the source-free approximation, introduces
our model system (the spin-polarized homogeneous electron gas), and gives an overview of linear response TD-SDFT and how to calculate spin waves with and
without the source-free construction.
Numerical results are presented and discussed in Sec. \ref{Sec:3}, and conclusions are given in Sec. \ref{Sec:4}.
Additional formal details and derivations are presented in three Appendices.

\section{Theoretical background} \label{Sec:2}

\subsection{Exchange-correlation fields in SDFT} \label{Sec:2A}

SDFT is concerned with interacting $N$-electron systems described by the many-body Hamiltonian
\begin{equation} \label{H}
\hat H = \sum_{j}^N\left[-\frac{\nabla_j^2}{2} + V(\bfr_j) + \bm{\sigma}_j\cdot \bfB(\bfr_j)  \right]
+
\frac{1}{2}\sum_{j\ne k}^N \frac{1}{|\bfr_j - \bfr_k|} ,
\end{equation}
where $V(\bfr)$ is a scalar potential, $\bfB(\bfr)$ is a (possibly noncollinear) magnetic field, and $\bm{\sigma}$ is the vector of Pauli matrices.
Here,  the Bohr magneton, $\mu_B = e \hbar/2m$, is absorbed in the definition of the magnetic field strength, and we
use atomic units ($e = m = \hbar = 4\pi \epsilon_0 = 1$) throughout.

The central idea of SDFT is that there exists a noninteracting system which reproduces the scalar density $n(\bfr)$ and the magnetization $\bfm(\bfr)$
of the interacting system in principle exactly \cite{Barth1972,Gunnarsson1976,Gidopoulos2007}. This noninteracting system is characterized by the Kohn-Sham equation
\begin{equation} \label{2}
\left[\left(-\frac{\nabla^2}{2} + V_{\rm KS}(\bfr) \right)I + \bm{\sigma} \cdot \bfB_{\rm KS}(\bfr) \right]
\Psi_i(\bfr) = \epsilon_i \Psi_i(\bfr) \:,
\end{equation}
where $I$ is the $2\times 2$ unit matrix and the Kohn-Sham orbitals $\Psi_i(\bfr)$ are two-component spinors.
The effective scalar potential and magnetic field are given by
\begin{eqnarray}
V_{\rm KS}(\bfr) &=& V(\bfr) + \int  \frac{n(\bfr')d\bfr'}{|\bfr-\bfr'|} + V_{\rm xc}[n,\bfm](\bfr),\\
\bfB_{\rm KS}(\bfr) &=& \bfB(\bfr) + \bfB_{\rm xc}[n,\bfm](\bfr).
\end{eqnarray}
The xc scalar potential and magnetic field are functionals of $n$ and $\bfm$ and are defined as functional derivatives of the xc energy:
\begin{equation}\label{der}
V_{\rm xc}(\bfr) = \frac{\delta E_{\rm xc}[n,\bfm]}{\delta n(\bfr)}\:, \qquad
\bfB_{\rm xc}(\bfr) = \frac{\delta E_{\rm xc}[n,\bfm]}{\delta \bfm(\bfr)}\:.
\end{equation}

To make SDFT work in practice, approximations are needed. Several recent studies have focused on $\bfB_{\rm xc}$ for
noncollinear magnetic systems, where local xc torques of the form $\bm \tau_{\rm xc}(\bfr) = \bfB_{\rm xc}(\bfr) \times \bfm(\bfr)$ can arise
\cite{Peralta2007,Eich2013,Scalmani2012,Bulik2013,Pittalis2017,Goings2018,Ullrich2018,Pluhar2019,Desmarais2021,Tancogne2023a,Tancogne2023b,Hill2023,Pu2023,Moore2024}.
Such xc torques are absent in the standard LSDA \cite{Kubler1988,Sandratskii1998}.

Capelle and Gross \cite{Capelle1997} showed that there is a close connection between the xc functionals of SDFT and of current-DFT (CDFT \cite{Vignale1987}).
For the special case of finite systems with vanishing external magnetic fields and orbital currents, they demonstrated that the exact xc magnetic field of SDFT is
source free (SF), i.e., a purely solenoidal vector field. Sharma {\em et al.} \cite{Sharma2018} later considered the space of densities $(n,\bfm)$ obtained from physical external
magnetic fields, $\bfB(\bfr) = \nabla \times {\bf A}(\bfr)$, and showed that the xc  energy functional can then be chosen to have the form $\tilde E_{\rm xc}[n,\nabla \times \bfm]$. If the functional derivative $\tilde \bfB_{\rm xc} = \delta \tilde E_{\rm xc}/\delta \bfm$ is constrained to remain within the space of $\bfm(\bfr)$ coming
only from physical magnetic fields, then the resulting $\tilde \bfB_{\rm xc}(\bfr)$ is SF. The proof of Ref. \cite{Sharma2018} relies on the assumption of finite system
size or lattice periodicity. The exact $\bfB_{\rm xc}$ of SDFT, on the other hand, is not subject to the above constraints and assumptions, and in general cannot be
expected to be SF. In other words, $\tilde \bfB_{\rm xc}(\bfr)$ is different from the exact functional $\bfB_{\rm xc}(\bfr)$.
However, the SF character of $\tilde \bfB_{\rm xc}(\bfr)$ is intuitively appealing and its practical consequences are worth exploring.
Furthermore, it can be shown  \cite{Sharma2018} that the SF $\tilde \bfB_{\rm xc}(\bfr)$ can in principle produce exact total magnetic moments.

Approximations to the xc magnetic field may not be SF, i.e., $\nabla \cdot \bfB^{\rm approx}_{\rm xc}(\bfr) \ne 0$.
Sharma {\em et al.} \cite{Sharma2018} proposed to construct a class of approximations where the SF condition is enforced.
This can be done explicitly using the Helmholtz construction:
\begin{equation}\label{B_SF}
\bfB_{\rm xc,SF}^{\rm approx}(\bfr) = \frac{s}{4\pi} \nabla \times \int \frac{\nabla' \times \bfB_{\rm xc}^{\rm approx}(\bfr')}{|\bfr - \bfr'|} d\bfr' \:.
\end{equation}
The construction (\ref{B_SF}) takes any approximated xc magnetic field as input and yields only the transverse (SF) part of it as output.
The dimensionless numerical parameter $s$ is an empirical scaling factor, which can be used to improve the performance of the functional. In Ref. \cite{Sharma2018}
it was demonstrated that choosing $s=1.12$ yields a good description of the magnetic properties of a range of materials, using the SF construction based on the LSDA;
in a similar manner, the SF construction based on LSDA+U was shown to be successful for describing strongly correlated magnetic materials \cite{Krishna2019}.
It was also shown  that the SF construction improves the convergence of noncollinear magnetic structures \cite{Moore2024} and
yields nonvanishing xc magnetic torques which affect the ultrafast spin dynamics induced by laser pulses \cite{Dewhurst2018}.

While this is quite promising, we here conduct a different test in the dynamical regime, namely, we calculate spin-wave dispersions in
spin-polarized homogeneous electron gases, for which exact results and properties are known. As we will see, this reveals some problematic behavior
of the SF construction.

\subsection{Spin-split homogeneous electron gas} \label{Sec:2B}

\begin{figure}[t]
\includegraphics[width=\columnwidth]{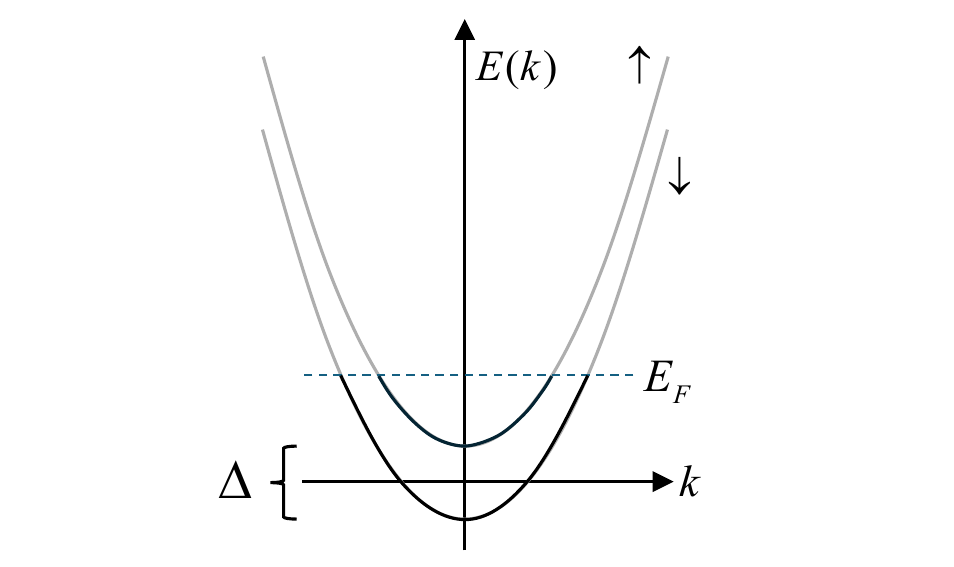}
\caption{\label{fig1} Parabolic energy dispersions of an HEG in the presence of a uniform magnetic field $B_{\rm KS}$ along the positive $z$-direction.
The spin splitting is given by $\Delta=2B_{\rm KS}$. The occupied states below the Fermi energy $E_F$ are indicated by the dark lines; light grey lines
indicate unoccupied states.}
\end{figure}

For a homogeneous electron gas (HEG) in a uniform effective magnetic field along the $z$-direction, $\bfB_{\rm KS}(\bfr) = B_{\rm KS}\hat e_z$, the single-particle states obey the equation:
\begin{equation}
    \left(-\frac{\hbar^2\nabla^2}{2m} + s_\sigma B_{\rm KS}\right)\psi_{\bfk\sigma}(\bfr) = \varepsilon_{\bfk\sigma}\psi_{\bfk\sigma}(\bfr),
\end{equation}
which has the solutions
\begin{equation}\label{8}
    \psi_{\bfk\sigma}(\bfr) = e^{i\bfk\cdot\bfr}   \qquad  \varepsilon_{\bfk\sigma} = \frac{k^2}{2} + s_\sigma B_{\rm KS} \:,
\end{equation}
where $s_\sigma=\pm 1$ for $\sigma = \ud$, respectively. Figure \ref{fig1} shows the energy dispersions: two parabolas, separated by
$\Delta = 2B_{\rm KS}$ and occupied up to the Fermi energy $E_F$.
We note that we here assume that the magnetic field only couples to the electronic spins, not to the orbital motion; in other words, we do not consider
effects related to Landau level quantization of the electron gas.

The spin-split HEG is characterized by the density $n$ and spin polarization $\zeta = (n_\ua - n_\da)/n$.
In 3D we find the following relations:
\begin{eqnarray}
E_F &=& \frac{1}{4} (3\pi^2 n)^{2/3} \left[ (1 - \zeta)^{2/3} + (1 + \zeta)^{2/3}\right],\\
B_{\rm KS}   &=& -\frac{1}{4} (3\pi^2 n)^{2/3} \left[ (1 + \zeta)^{2/3} - (1 - \zeta)^{2/3}\right].
\end{eqnarray}
From this, the spin-resolved Fermi wavevectors follow as
\begin{equation}
k_{F\sigma} = (3\pi^2 n)^{1/3}(1+ s_\sigma \zeta)^{1/3} \:.
\end{equation}
The corresponding relations in 2D are
\begin{eqnarray}
E_F &=& \pi n,\\
B_{\rm KS}   &=& -\pi n \zeta,\\
k_{F\sigma} &=& (2\pi n)^{1/2}(1+ s_\sigma \zeta)^{1/2} \:.
\end{eqnarray}
A useful quantity to characterize the HEG is the Wigner-Seitz radius, defined as $r_s = (3/4\pi n)^{1/3}$ in 3D and $r_s = (1/\pi n)^{1/2}$ in 2D.
The definitions of $r_s$ are independent of the spin polarization.

\subsection{Linear spin-density matrix response}\label{Sec:2C}

In the following it will be convenient to use an alternative formulation of SDFT based on the spin-density matrix $\uun$ instead
of the $(n,\bfm)$-based formulation of Sec. \ref{Sec:2A}. Details of the transformation between the two formulations are given in Appendix \ref{App:A}.

The linear response equation of the spin-density matrix has the individual components
\begin{equation}\label{response_eq}
    \delta n_{\sigma\sigma'}(\bfr,\omega) = \sum_{\tau\tau'}\int d\bfr'\chi_{\sigma\sigma',\tau\tau'}(\bfr,\bfr',\omega) \delta v_{\tau\tau'}^{\rm KS}(\bfr',\omega),
\end{equation}
where $\delta v_{\tau\tau'}^{\rm KS}(\bfr',\omega)$ is the effective spin-dependent perturbing potential (see below) and $\chi_{\sigma\sigma',\tau\tau'}(\bfr,\bfr',\omega)$ is the noninteracting response function, defined as
\begin{eqnarray}
\lefteqn{
    \chi_{\sigma\sigma',\tau\tau'}(\bfr,\bfr',\omega) =}\nonumber\\
&& \sum_{\mu\nu}^{\infty}(f_\mu-f_\nu) \frac{\psi_{\mu\sigma}(\bfr)\psi_{\nu\sigma'}^*(\bfr)\psi_{\mu\tau}^*(\bfr')\psi_{\nu\tau'}(\bfr')}{\omega-\varepsilon_\mu+\varepsilon_\nu+i\eta} \:.
\end{eqnarray}
Here, $f_\mu$ is the zero-temperature occupation number of the $\mu$th eigenstate (0 if empty, 1 if occupied), and $\eta$ is a positive infinitesimal.

For spin waves we need the spin-flip response functions, $\chi_{\uparrow\downarrow,\uparrow\downarrow}$ and $\chi_{\downarrow\uparrow,\downarrow\uparrow}$.
For a spin-split HEG, these are given by

\begin{equation}\label{chi_ud}
    \begin{aligned}
        \chi_{\ud,\ud}(q,\omega) = &- \int \frac{d\bfk}{(2\pi)^{\tt d}}\frac{f(\varepsilon_{\bfk\ua})}{\omega - \bfk\cdot \bfq + \frac{q^2}{2} -\Delta + i\eta} \\
        &+ \int \frac{d\bfk}{(2\pi)^{\tt d}}\frac{f(\varepsilon_{\bfk\da})}{\omega - \bfk\cdot \bfq - \frac{q^2}{2} -\Delta + i\eta} \\
    \end{aligned}
\end{equation}
\begin{equation}\label{chi_du}
    \begin{aligned}
        \chi_{\du,\du}(q,\omega) = &- \int \frac{d\bfk}{(2\pi)^{\tt d}}\frac{f(\varepsilon_{\bfk\da})}{\omega - \bfk\cdot \bfq + \frac{q^2}{2} +\Delta + i\eta} \\
        &+ \int \frac{d\bfk}{(2\pi)^{\tt d}}\frac{f(\varepsilon_{\bfk\ua})}{\omega - \bfk\cdot \bfq - \frac{q^2}{2} +\Delta + i\eta} \\
    \end{aligned}
\end{equation}
where ${\tt d}=2$ and 3 in 2D and 3D, respectively, and $f(\varepsilon_{\bfk\sigma})$ is the zero-temperature Fermi function with respect to $E_F$ and using the
single-particle energies $\varepsilon_{\bfk\sigma}$ of Eq. (\ref{8}).

The noninteracting spin-flip response functions (\ref{chi_ud}) and (\ref{chi_du}) can be evaluated analytically for the real and imaginary part,
similarly to the derivation of the spin-conserving response functions of the HEG \cite{GiulianiVignale}. The results are presented in Appendix \ref{App:B}.

The frequency- and spin-dependent perturbing KS potential is in general given by
\begin{equation}
\delta v_{\tau\tau'}^{\rm KS}(\bfr,\omega) =
\delta v_{\tau\tau'}^{\rm ext}(\bfr,\omega)  + \delta v_{\tau\tau'}^{\rm H}(\bfr,\omega)  + \delta v_{\tau\tau'}^{\rm xc}(\bfr,\omega) \:,
\end{equation}
where $\delta v_{\tau\tau'}^{\rm ext}$ is the external perturbation, the linearized Hartree potential is given by
\begin{equation}\label{Hartree}
\delta v_{\tau\tau'}^{\rm H}(\bfr,\omega) = \sum_{\lambda\lambda'}\int d\bfr' \: f^{\rm H}_{\tau\tau',\lambda\lambda'}(\bfr,\bfr')\delta n_{\lambda\lambda'}(\bfr',\omega) \:,
\end{equation}
with the Hartree kernel $f^{\rm H}_{\tau\tau',\lambda\lambda'}(\bfr,\bfr') = |\bfr-\bfr'|^{-1} \delta_{\tau\tau'}\delta_{\lambda\lambda'}$,
and the linearized xc potential is defined in Appendix \ref{App:A2}, featuring the xc kernel $f^{\rm xc}_{\sigma\sigma',\tau\tau'}(\bfr,\bfr',\omega)$.

\subsection{Spin Waves in a spin-polarized HEG}

The excitation spectrum of any system is characterized by the condition that the response equation (\ref{response_eq}) has
finite solutions in the absence of any external perturbation, i.e., $\delta v_{\tau\tau'}^{\rm ext}(\bfr,\omega)=0$.
This condition can be written in a form in which an integral operator consisting of $\chi_{\sigma\sigma',\tau\tau'}$ and the Hartree and xc kernels has the
eigenvalue 1 \cite{Petersilka1996}.

For the HEG, this reduces to the condition in which the matrix $(\underline{\underline \chi} \,\underline{\underline f}^{\rm xc})$ has eigenvalue 1.
For the xc matrix elements we use Eqs. (\ref{f1})--(\ref{f4}) for the HEG, and Eq. (\ref{fxc_sf_mat}) if the source-free correction is included.
Both cases show that in a spin-polarized (para- or ferromagnetic) HEG the spin-conserving and spin-flip excitation channels are decoupled and can be considered
independently of each other \cite{GiulianiVignale}.
The spin-conserving channel includes the plasmon modes, which are dominated by classical Coulomb effects, i.e., the linearized
Hartree potential (\ref{Hartree}). The spin-flip excitations, on the other hand, are solely determined by linearized xc effects.

Thus, to calculate the spin waves we can consider the $2\times 2$ matrix product
\begin{equation}\label{kernelmatrix}
    \begin{pmatrix}
        f^{\text{xc}}_{\ua\da,\ua\da} & f^{\text{xc}}_{\uparrow\downarrow,\downarrow\uparrow} \\
        f^{\text{xc}}_{\downarrow\uparrow,\uparrow\downarrow} & f^{\text{xc}}_{\da\ua,\da\ua}
    \end{pmatrix}
    \begin{pmatrix}
    \chi_{\uparrow\downarrow,\uparrow\downarrow} & 0 \\
    0 & \chi_{\downarrow\uparrow,\downarrow\uparrow}
    \end{pmatrix},
\end{equation}
which leads to the condition
\begin{equation}\label{eigenvalue}
\mbox{det} \left| \begin{array}{cc}
f^{\rm xc}_{\ua\da,\ua\da}\chi_{\ua\da,\ua\da}-1 & f^{\rm xc}_{\ua\da,\da\ua}\chi_{\da\ua,\da\ua} \\
f^{\rm xc}_{\da\ua,\ua\da}\chi_{\ua\da,\ua\da} & f^{\rm xc}_{\da\ua,\da\ua}\chi_{\da\ua,\da\ua}-1
\end{array}\right| = 0\:.
\end{equation}
The xc kernel of the HEG in LSDA (without the SF correction) is given by
\begin{eqnarray}
f^{\rm xc}_{\ua\da,\ua\da} = f^{\rm xc}_{\da\ua,\da\ua} &=& 2 h^{\rm xc}_{11} \\
f^{\rm xc}_{\ua\da,\da\ua} = f^{\rm xc*}_{\da\ua,\ua\da} &=&  0 \:,
\end{eqnarray}
where $h^{\rm xc}_{11} = \frac{1}{n\zeta} \frac{\partial \exc}{\partial \zeta}$, and in the following we will abbreviate $\partial e_{\rm xc}/\partial \zeta = e'_{\rm xc}$.
Equation (\ref{eigenvalue}) then gives the following condition for spin waves:
\begin{equation}\label{condition_no_sf}
    \bigl(f^{\text{xc}}_{\uparrow\downarrow,\uparrow\downarrow}\chi_{\uparrow\downarrow,\uparrow\downarrow} -1\bigr)\bigl(f^{\text{xc}}_{\downarrow\uparrow,\downarrow\uparrow}\chi_{\downarrow\uparrow,\downarrow\uparrow} -1\bigr)=0 \:.
\end{equation}
This can be used to determine the spin-wave dispersion $\omega(q)$ numerically.

Using condition (\ref{condition_no_sf})  and  the analytic expressions for the spin-flip Lindhard function, we can find small-$q$ expressions for the LSDA
spin wave dispersion
\cite{Yosida}:
\begin{equation}\label{omega3D}
    \omega^{\rm (3D)}(q) = 2 B_{\text{ext}} -\frac{q^2}{2\zeta}\left[1-  \frac{k_{F\downarrow}^5 - k_{F\uparrow}^5}{30\pi^2 n e'_{\text{xc}}}\right] + \mathcal{O}(q^3)
\end{equation}
\begin{equation}\label{omega2D}
    \omega^{\rm (2D)}(q) = 2 B_{\text{ext}} -\frac{q^2}{2\zeta}\left[1 + \frac{\pi n \zeta}{e'_{\text{xc}}}\right] + \mathcal{O}(q^3).
\end{equation}

Larmor's Theorem \cite{Lipparini,Karimi2018} states that in a system of identical particles with sufficiently weak magnetic field $B_{\text{ext}}$, the particles carry out a collective precession at exactly the noninteracting frequency, which is determined only by $B_{\text{ext}}$. For the HEG, this means $\omega(q=0) = 2 B_{\text{ext}}$. As shown by the small-$q$ dispersion, this is true within LSDA in both 2D  and 3D.

If the SF correction is applied on top of the LSDA, see Appendix \ref{App:C}, then the ground state of the HEG remains fundamentally unchanged
since a uniform magnetic field is per definition source free. However, the linear response of the system does {\em not} remain unchanged: the xc kernels now become
\begin{eqnarray}
f^{\rm xc}_{\ua\da,\ua\da} = f^{\rm xc}_{\da\ua,\da\ua} &=& h^{\rm xc}_{11} \\
f^{\rm xc}_{\ua\da,\da\ua} = f^{\rm xc*}_{\da\ua,\ua\da} &=&  \frac{(q_1+iq_2)^2}{q^2}h^{\rm xc}_{11} \:.
\end{eqnarray}
Equation (\ref{eigenvalue}) then gives a new condition:
\begin{eqnarray}
 0 &=&  \bigl(f^{\text{xc}}_{\uparrow\downarrow,\uparrow\downarrow}\chi_{\uparrow\downarrow,\uparrow\downarrow} -1\bigr)\bigl(f^{\text{xc}}_{\downarrow\uparrow,\downarrow\uparrow}\chi_{\downarrow\uparrow,\downarrow\uparrow} -1\bigr)\nonumber\\
&-&
f^{\text{xc}}_{\uparrow\downarrow,\downarrow\uparrow}f^{\text{xc}}_{\downarrow\uparrow,\uparrow\downarrow}
\chi_{\uparrow\downarrow,\ua\da}\chi_{\downarrow\uparrow,\downarrow\uparrow} \:.
\end{eqnarray}
After simplification, this becomes
\begin{equation}\label{sf condition}
    \frac{e'_{\rm xc}}{n\zeta} \bigl(\chi_{\uparrow\downarrow,\uparrow\downarrow} + \chi_{\downarrow\uparrow,\downarrow\uparrow}\bigr)-1=0.
\end{equation}
Using Eq. (\ref{sf condition}) and the analytic expressions for the spin-flip Lindhard function, we can find small-$q$ expressions for the SF-LSDA spin wave dispersion:
\begin{equation}\label{omega3D_SF}
    \begin{aligned}
        \omega^{\rm (3D)}_{\rm SF}(q) = \sqrt{\Delta^2 - 2\Delta se_{\text{xc}}'} - \frac{(1-se_{\text{xc}}'/\Delta)q^2}{2\zeta\sqrt{1-2se_{\text{xc}}'/\Delta}} \\
        - \frac{(2-3se_{\text{xc}}'/\Delta)k_F^5[(1+\zeta)^{5/3} - (1-\zeta)^{5/3}]q^2}{60\pi^2n\zeta se'_{\text{xc}}\sqrt{1-2se'_{\text{xc}}/\Delta}} + \mathcal{O}(q^3)
    \end{aligned}
\end{equation}
\begin{equation}\label{omega2D_SF}
    \begin{aligned}
            \omega^{\rm (2D)}_{\rm SF}(q) = \sqrt{\Delta^2 - 2\Delta se_{\text{xc}}'} - \frac{(1-se_{\text{xc}}'/\Delta)q^2}{2\zeta\sqrt{1-2se_{\text{xc}'}/\Delta}} \\
        - \frac{(2-3se_{\text{xc}}'/\Delta)\pi n q^2}{2 se'_{\text{xc}}\sqrt{1-2se'_{\text{xc}}/\Delta}} + \mathcal{O}(q^3).
    \end{aligned}
\end{equation}
We now see that Larmor's Theorem is violated if the SF correction is made:
using $\Delta = 2(B_{\rm ext} + sB_{\rm xc})$, we obtain in both 2D and 3D,
\begin{equation}\label{omegaSF}
\omega_{\rm SF}(q=0) = 2\sqrt{B_{\rm ext}^2 + s B_{\rm ext} B_{\rm xc}} \ne 2 B_{\rm ext} \:.
\end{equation}
We will investigate the severity of this violation below.

\begin{figure}
\includegraphics[width=\columnwidth]{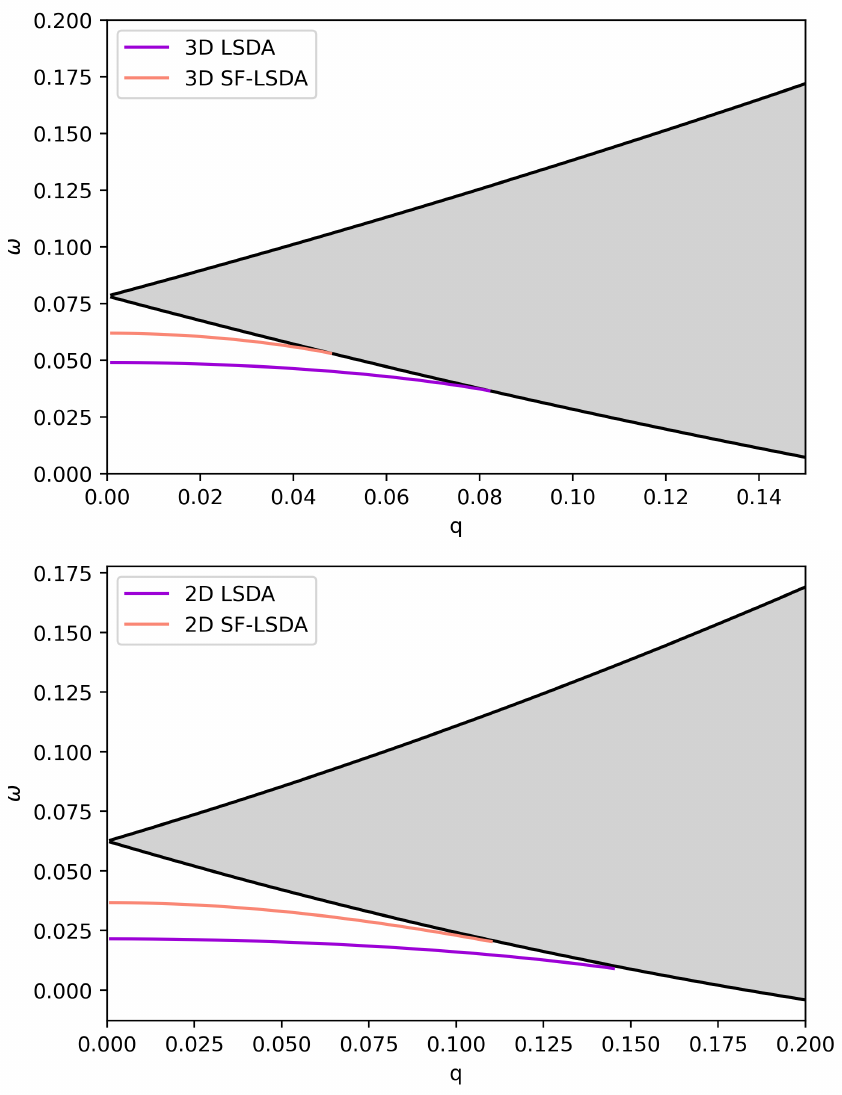}
\caption{\label{fig2} Spin-wave dispersions of a paramagnetic HEG in 3D (top) and 2D (bottom), with $r_s=4$ and $\zeta=-0.5$ (purple: LSDA, orange: SF-LSDA).
The shaded areas indicate the regions of single-particle spin-flip excitations.  }
\end{figure}

\section{Numerical Results and Discussion}\label{Sec:3}

\subsection{Paramagnetic HEG}
We first consider the paramagnetic case, where a uniform applied magnetic field $B_{\rm ext}$ is required to induce the spin polarization $\zeta$.

\begin{figure}
\includegraphics[width=\columnwidth]{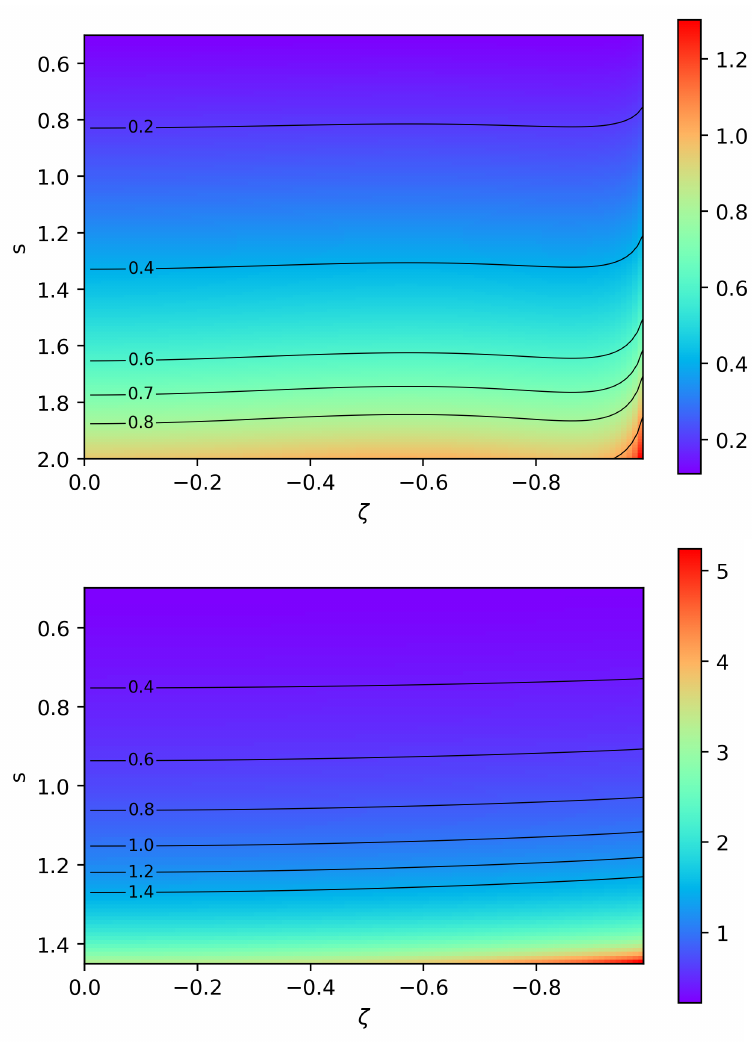}
\caption{\label{fig3} Relative violation of Larmor's Theorem, quantified by $\Gamma$  [Eq. (\ref{Gamma})], for a 3D (top) and 2D (bottom) HEG with
$r_s=4$, as a function of spin polarization $\zeta$ and empirical SF scaling factor $s$. Contour lines indicate constant values of $\Gamma$.}
\end{figure}

\subsubsection{Spin-wave dispersions and Larmor's theorem}

Using the conditions for LSDA spin waves and SF-LSDA spin waves, as well as the spin-flip Lindhard functions, we can solve numerically for the frequency and obtain spin wave dispersions $\omega(q)$ in 2D and 3D. Examples are shown in Fig. \ref{fig2}, where we choose $r_s=4$ and $\zeta=-0.5$ in both cases.
In 3D this implies $B_{\rm ext} = 0.0245$ and in 2D it implies $B_{\rm ext} = 0.0108$.

The shaded areas shown in Fig. \ref{fig2} are the regions of single-particle spin-flip (Stoner) excitations. The boundaries of these regions follow from the poles of Eqs. (\ref{chi_ud}) or (\ref{chi_du}) for positive or negative polarization values, respectively. Here we have $\zeta<0$, and so the upper ($+$) and lower ($-$) boundaries
are given by
\begin{equation}
    \omega_{\pm}(q) = \frac{q^2}{2} \pm \frac{k_{F\downarrow}q}{2} + \Delta  \:.
\end{equation}

The energy dispersions $\omega(q)$ of the spin waves are found below the single-particle continuum. As discussed above, the SF spin wave dispersions
violate the Larmor condition $\omega(q=0)=2B_{\rm ext}$, and, as seen from Fig. \ref{fig2}, the discrepancy is substantial: the SF spin wave dispersions in 3D and 2D
have similar shapes as the LSDA ones, but are significantly offset to higher frequencies.

To further quantify this observation, we introduce a parameter $\Gamma$, which measures the relative violation of Larmor's Theorem:
\begin{equation}\label{Gamma}
    \Gamma = \frac{\omega_{\rm SF}(0) - 2B_{\text{ext}}}{2B_{\text{ext}}},
\end{equation}
where $\omega_{\rm SF}(0)$ is given in Eq. (\ref{omegaSF}).
$\Gamma$ depends on the HEG parameters $r_s$ and $\zeta$, as well as on the empirical scaling factor $s$.
The results are shown in Fig. \ref{fig3}, for $r_s$=4 and $0<\zeta<1$ and a range of values of $s$.

Generally, a mild violation of Larmor's Theorem would correspond to values of order $\Gamma \lesssim 0.1$.
For $s$ around 1.1, we find that $\Gamma\approx 0.3$ in 3D and $\Gamma\approx 0.8$ in 2D, which is certainly not small.
We also calculated $\Gamma$ for other values of $r_s$, and found the general trend that the degree of violation of Larmor's theorem
increases with $r_s$.

\begin{figure*}
\includegraphics[width=\linewidth]{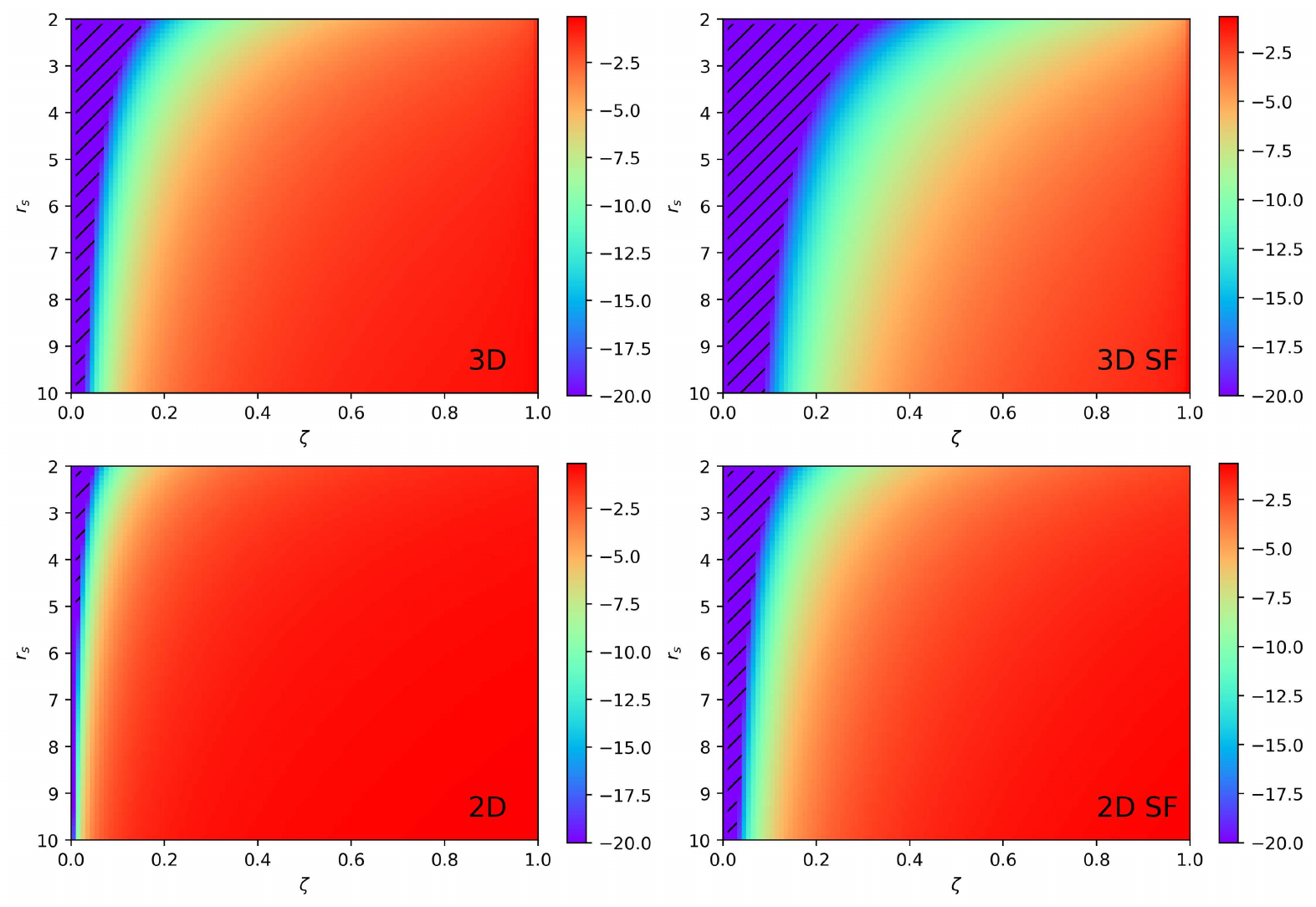}
\caption{\label{fig4} Spin-wave stiffness $S$ of the 3D (top) and 2D (bottom) HEG, calculated in LSDA with and without the SF correction (here, $s=1$).
In the striped purple regions, $S$ takes on values below $-20$.}
\end{figure*}

\subsubsection{Spin-wave stiffness}

A useful quantity to characterize spin-wave dispersions is the spin-wave stiffness $S$, which is defined via the small-$q$ dispersion:
\begin{equation}
    \omega(q\to 0) = \omega_0 + \frac{1}{2}Sq^2 \:.
\end{equation}
In other words, $S$ is a measure of the curvature of the small-$q$ limit of the spin-wave dispersion; $S$ can be read off from Eqs. (\ref{omega3D}), (\ref{omega2D})
and (\ref{omega3D_SF}), (\ref{omega2D_SF}), respectively.

In Fig. \ref{fig4} we compare the spin-wave stiffnesses in 2D and 3D within LSDA and SF-LSDA, where we assume $s=1$ in the SF case.
The overall behavior of $S$ as a function of $r_s$ and $\zeta$ is similar with and without the SF correction; however,
for given values of $r_s$ and $\zeta$, the SF-LSDA stiffness tends to be greater (i.e., more negative) than the LSDA stiffness by roughly a
factor of 2--3 (in 3D) and 2.5--4 (in 2D). The reason for this is easily seen from the spin-wave dispersion curves in Fig. \ref{fig2}: the SF dispersions are
always above the non-SF dispersions, and therefore reach the spin-flip continuum earlier. This causes the SF dispersions to bend down more rapidly,
thus enhancing $S$. Similar trends are observed for scaling factors $s$ different from 1.

\subsection{Ferromagnetic HEG}

The phase diagram of the ground state of the HEG is quite rich \cite{GiulianiVignale}. It is know that the ground state undergoes a transition from
paramagnetic to ferromagnetic for decreasing density, eventually leading to Wigner crystallization.
One finds that the paramagnetic to ferromagnetic transition occurs in 3D at $r_s=73.0$ (using the xc functional of Ref. \cite{Perdew1992})
and in 2D at $r_s=25.6$ (using the xc functional of Ref. \cite{Attaccalite2002}). If the correlation energy of the HEG is not included, i.e., in exchange-only,
then the HEG becomes ferromagnetic at much smaller values of $r_s$ (in 3D at $r_s=5.45$ and in 2D at $r_s=2.01$).

In the ferromagnetic case, the magnon dispersion of the HEG follows from Eqs. (\ref{omega3D}) and (\ref{omega2D}) by setting $\zeta=-1$
and $B_{\rm ext}=0$:
\begin{equation}\label{omega3D_FM}
    \omega^{\rm (3D)}(q) =  \frac{q^2}{2}\left[1-  \frac{(6\pi^2 n)^{2/3}}{5 e'_{\text{xc}}}\right] + \mathcal{O}(q^3)
\end{equation}
\begin{equation}\label{omega2D_FM}
    \omega^{\rm (2D)}(q) =  \frac{q^2}{2}\left[1 - \frac{\pi n }{e'_{\text{xc}}}\right] + \mathcal{O}(q^3).
\end{equation}
We thus obtain a gapless dispersion with $\sim q^2$ behavior at small $q$, which is characteristic for ferromagnetic magnons.

A very different behavior is observed in the SF case. Setting $\zeta=-1$ and $B_{\rm ext}=0$ we find
\begin{equation}
\omega_{\rm SF}^{\rm (3D)} =  q\left[s e'_{\rm xc} - \frac{ (6\pi^2 n)^{2/3}}{5}\right]^{1/2}  + {\cal O}(q^2)
\end{equation}
\begin{equation}
\omega_{\rm SF}^{\rm (2D)} = q\left[s e_{\rm xc}'  - \pi n \right]^{1/2}   + {\cal O}(q^2).
\end{equation}
These are gapless mode dispersions linear in $q$, which is a behavior one would find in an antiferromagnet, but certainly not in the ferromagnetic HEG.

\begin{figure}
\includegraphics[width=\columnwidth]{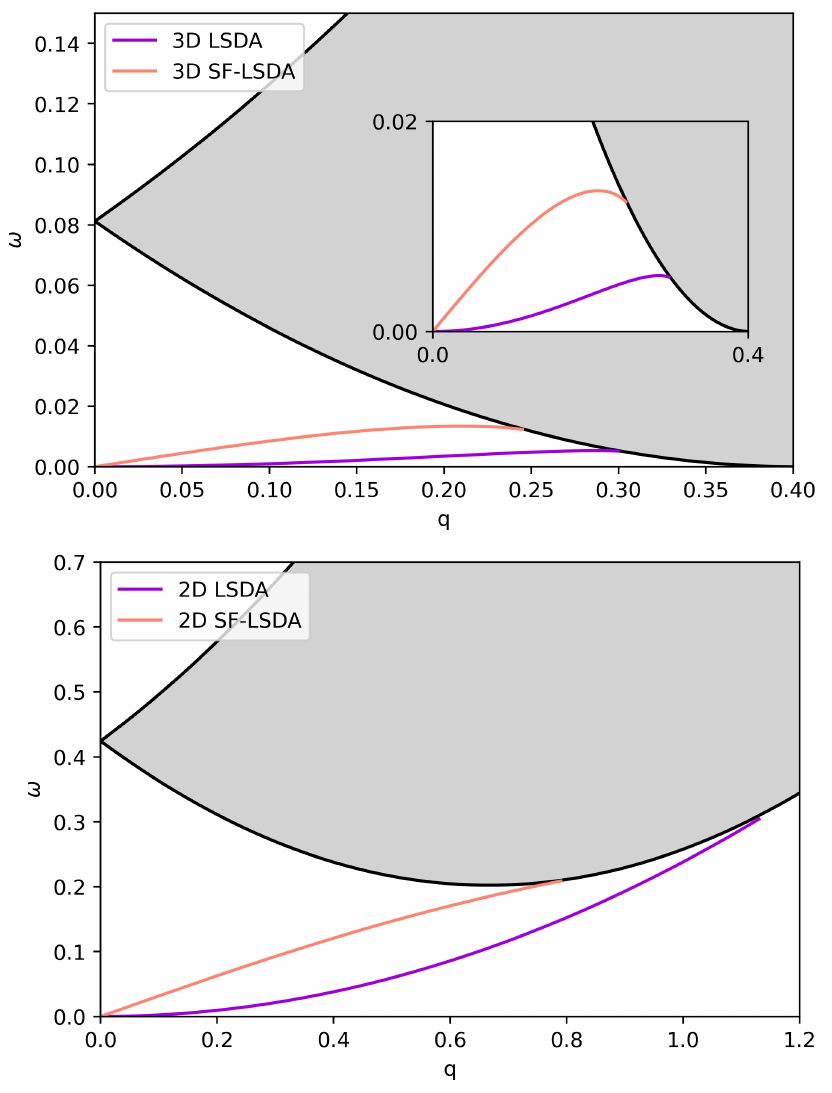}
\caption{\label{fig5} Magnon dispersions for a ferromagnetic 3D (top) and 2D (bottom) HEG. The SF-LSDA dispersions have an incorrect (linear in $q$) behavior.
The calculations were done using exchange only, for $r_s=6$ (3D) and $r_s=3$ (2D).}
\end{figure}

Figure \ref{fig5} gives an explicit illustration, using exchange-only LSDA and SF-LSDA (since in exchange-only the HEG turns ferromagnetic
for reasonably small values of $r_s$). The spin waves all start correctly at $\omega=0$ for $q=0$ but then
approach the Stoner continuum in a different manner; the linear behavior of SF-LSDA is clearly visible.

\section{Conclusions}\label{Sec:4}

Under certain constraints and assumptions,
the xc magnetic field of SDFT can be expressed as $\bfB_{\rm xc}(\bfr) = \nabla \times {\bf A}_{\rm xc}(\bfr)$, i.e., it can be chosen to be purely transverse.
In practice, this condition does not hold for most commonly used approximations, and it is an attractive idea to enforce it by construction. This defines the
SF condition, see Eq. (\ref{B_SF}). The literature suggests \cite{Sharma2018,Krishna2019,Moore2024} that this indeed leads to improved results for the ground state
of magnetic materials.

The drawback of the SF construction is that it is an ad-hoc prescription that is directly applied as a fix for a given approximation to $\bfB_{\rm xc}$.
By contrast, the exact xc magnetic field is derived as the (unconstrained) functional derivative of the xc energy functional $E_{\rm xc}[n,\bfm]$.
This is not the case for the SF-corrected $\bfB_{\rm xc}$:
in other words, it is not a functional derivative. While this appears to be relatively benign for the ground state, we have found it to cause serious problems
if the SF construction is extended into the dynamical regime.

We have tested the SF prescription for one of the most important model systems in condensed matter, the HEG. We calculated spin-wave dispersions of the paramagnetic
and ferromagnetic HEG for a broad range of densities and spin polarizations. While SF was capable of producing spin waves, it led to the violation of two
exact conditions: Larmor's theorem for the paramagnetic case, and the correct (quadratic) small-$q$ behavior of the magnons in the ferromagnetic case.
While the HEG is admittedly an idealized model system, it nevertheless is very relevant for real materials. Specifically, any magnetic material
which has a significant contribution to its magnetism coming from itinerant electrons requires a method that correctly describes the physics in the HEG limit.

There does not seem to be an easy way to fix these defects of the SF construction in the dynamical regime. Our study suggests that the SF construction
in its present form should be limited to the magnetic ground state or to magnetic materials with localized magnetic moments. Magnetic excitations in materials with
significant itinerant character should better be described using standard local or semilocal functionals. Finding an xc magnetic field functional that
is source-free and yields the correct physics in both the ground state and in the dynamical regime will remain an important task for future studies.


\acknowledgments

This work was supported by DOE Grant No. DE-SC0019109. We thank Sangeeta Sharma and Kay Dew\-hurst for very helpful comments.

\appendix

\section{Variables and transformations in SDFT}\label{App:A}

\subsection{Densities and Kohn-Sham potentials}\label{App:A1}

In Section \ref{Sec:2A} we formulated SDFT using the density $n$ and magnetization $\bfm$ as basic variables.
Alternatively \cite{Ullrich2018}, SDFT can be formulated in terms of the spin-density matrix
\begin{equation}
\underline{\underline n}(\bfr)
= \left( \begin{array}{cc} n_\uu(\bfr) & n_\ud(\bfr)\\ n_\du(\bfr) & n_\dd(\bfr) \end{array} \right) .
\end{equation}
For the Kohn-Sham system, the spin-density matrix is defined as $\underline{\underline n}(\bfr) = \sum_{i=1}^N \Psi_i(\bfr) \Psi_i^\dagger(\bfr)$,
where the spin-up and spin-down components of $\Psi_i(\bfr)$ follow from
\begin{equation}
\sum_{\beta }
\left[ - \frac{\nabla^2}{2}\,\delta_{\alpha\beta} +
v^{\rm KS}_{\alpha \beta}({\bf r})  \right]
\psi_{i\beta}({\bf r}) = \epsilon_i \psi_{i\alpha}({\bf r}) \:.
\end{equation}
Here, $\alpha$ and $\beta$ are spin indices running over $\ua,\da$.
The effective Kohn-Sham potential $v^{\rm KS}_{\alpha \beta}({\bf r})$ is a $2\times 2$ matrix in spin space whose xc part is defined as
\begin{equation}
v^{\rm xc}_{\alpha\beta}(\bfr) = \frac{\delta E_{\rm xc}[\,\uun\,]}{\delta n_{\beta\alpha}(\bfr)} \:.
\end{equation}
The SDFT formulations in terms of $(n,\bfm)$ and in terms of $\uun$ are physically equivalent.
To transform between them, it is convenient to rearrange the basic variables as 4-component column vectors,
\begin{equation}
\vec m(\bfr) = \left( \begin{array}{c} n(\bfr) \\ m_x(\bfr) \\ m_y(\bfr) \\ m_z(\bfr) \end{array} \right), \qquad
\vec n(\bfr) = \left( \begin{array}{c} n_\uu(\bfr) \\ n_\du(\bfr) \\ n_\ud(\bfr) \\ n_\dd(\bfr) \end{array} \right).
\end{equation}
One finds \cite{Ullrich2018}
\begin{equation}
\vec m (\bfr) = \uuT \: \vec n(\bfr), \qquad \vec n(\bfr)= \uuT^{-1} \vec m (\bfr)\:,
\end{equation}
where the transformation matrix is given by
\begin{equation}
\uuT = \left( \begin{array}{cccc}
1 & 0 & 0 & 1 \\
0 & 1 & 1 & 0 \\
0 & -i & i & 0 \\
1 & 0 & 0 & -1
\end{array} \right),
\end{equation}
with $2\uuT^{-1} = \uuT^\dagger$. Likewise, we can write the corresponding Kohn-Sham effective potentials and magnetic fields in 4-component vector form as
\begin{equation}
\vec V_{\rm KS}(\bfr) = \left( \begin{array}{c} V_{\rm KS}(\bfr) \\ B_{{\rm KS},x}(\bfr) \\ B_{{\rm KS},y}(\bfr) \\B_{{\rm KS},z}(\bfr) \end{array} \right), \qquad
\vec v_{\rm KS}(\bfr) = \left( \begin{array}{c} v^{\rm KS}_\uu(\bfr) \\ v^{\rm KS}_\du(\bfr) \\ v^{\rm KS}_\ud(\bfr) \\ v^{\rm KS}_\dd(\bfr) \end{array} \right).
\end{equation}
The connection between the two is
\begin{equation}
\vec V_{\rm KS}(\bfr) = \frac{1}{2}\:\uuT \, \vec v_{\rm KS}(\bfr) \:, \qquad
\vec v_{\rm KS}(\bfr) = 2 \uuT^{-1} \vec V_{\rm KS}(\bfr) \:.
\end{equation}

\subsection{Linear response}\label{App:A2}
We can formulate the frequency-dependent linear response equations within both the formulations discussed above.
The density-magnetization response to a perturbing scalar potential and magnetic field $\delta \vec V$ is given by
\begin{equation}
\delta \vec m(\bfr,\omega) =  \int d\bfr' \underline{\underline \Pi}(\bfr,\bfr',\omega) \, \delta \vec V(\bfr',\omega) \:,
\end{equation}
and the spin-density-matrix response to a perturbation $\delta v$ is given by
\begin{equation}
\delta \vec n(\bfr,\omega) =  \int d\bfr' \underline{\underline \chi}(\bfr,\bfr',\omega) \, \delta \vec v(\bfr',\omega) \:.
\end{equation}
The connection between the response functions is
\begin{equation}
\underline{\underline \Pi} = 2 \uuT \, \underline{\underline \chi} \, \uuT^{-1}  \:, \qquad
\underline{\underline \chi} = \frac{1}{2} \uuT^{-1} \, \underline{\underline \Pi} \, \uuT  \:.
\end{equation}
The same relations apply for the respective interacting and noninteracting response functions.

The key quantities in TD-SDFT are the linearized xc potentials, defined as follows:
\begin{eqnarray}
\delta \vec V_{\rm xc}(\bfr,\omega) &=&  \int d\bfr' \underline{\underline h}^{\rm xc}(\bfr,\bfr',\omega) \, \delta \vec m(\bfr',\omega)\\
\delta \vec v_{\rm xc}(\bfr,\omega) &=&  \int d\bfr' \underline{\underline f}^{\rm xc}(\bfr,\bfr',\omega) \, \delta \vec n(\bfr',\omega) \:,
\end{eqnarray}
where the xc kernel matrices are related via
\begin{equation}\label{kerneltrafo}
\underline{\underline h}^{\rm xc} = \frac{1}{2} \uuT \, \underline{\underline f}^{\rm xc} \, \uuT^{-1}  \:, \qquad
\underline{\underline f}^{\rm xc} =  2\uuT^{-1} \, \underline{\underline h}^{\rm xc} \, \uuT  \:.
\end{equation}
In the adiabatic approximation, the xc kernels are obtained from the xc energy functionals in the usual way. We have
\begin{equation}
h^{\rm xc}_{ij}(\bfr,\bfr') =  \frac{\delta^2 E_{\rm xc}[n,\bfm]}{\delta m_i(\bfr)\delta m_j(\bfr')}  \:,
\end{equation}
where the right-hand side is evaluated at the ground-state density and magnetization $(n,\bfm)$,
and the indices $i,j=0,1,2,3$ relate to $m_0 = n$, $m_{1,2,3} = m_{x,y,z}$.

Likewise,
\begin{equation}
f^{\rm xc}_{\alpha\beta,\sigma\tau}(\bfr,\bfr') =  \frac{\delta^2 E_{\rm xc}[\,\uun\,]}{\delta n_{\beta\alpha}(\bfr)\delta n_{\tau\sigma}(\bfr')} \:,
\end{equation}
where the right-hand side is evaluated at the ground-state spin-density matrix $\uun$.

For a HEG with uniform density $n$ and spin polarization $\zeta$, the only nonvanishing xc kernels are
\begin{equation}
h^{\rm xc}_{00} = 2 \frac{\partial e_{\rm xc}}{\partial n} + n \frac{\partial^2 e_{\rm xc}}{\partial n^2}
- 2\zeta \frac{\partial^2 e_{\rm xc}}{\partial n \partial \zeta} + \frac{\zeta^2}{n}\frac{\partial^2 e_{\rm xc}}{\partial n^2}
\end{equation}
\begin{eqnarray}
h^{\rm xc}_{03} = h^{\rm xc}_{30} &=&
\frac{\partial^2 e_{\rm xc}}{\partial n \partial \zeta} - \frac{\zeta}{n} \frac{\partial^2 e_{\rm xc}}{\partial \zeta^2}
\\
h^{\rm xc}_{11} = h^{\rm xc}_{22} &=& \frac{1}{n\zeta} \frac{\partial e_{\rm xc}}{\partial \zeta}
\\
h^{\rm xc}_{33} &=& \frac{1}{n} \frac{\partial^2 e_{\rm xc}}{\partial \zeta^2}
\end{eqnarray}
and correspondingly
\begin{eqnarray}
f^{\rm xc}_{\uparrow \uparrow,\uparrow \uparrow}
&=& 2\: \frac{\partial e_{\rm xc}}{\partial n}
+ n\: \frac{\partial^2 e_{\rm xc}}{\partial n^2}
+2 (1-\zeta) \frac{\partial^2 e_{\rm xc}}{\partial n \partial \zeta} \nonumber\\
&&{}
+ \frac{1}{n}(1-\zeta)^2 \frac{\partial^2 e_{\rm xc}}{\partial \zeta^2} \label{f1}
\\
f^{\rm xc}_{\downarrow \downarrow,\downarrow \downarrow}
&=& 2\: \frac{\partial e_{\rm xc}}{\partial n}
+ n\: \frac{\partial^2 e_{\rm xc}}{\partial n^2}
-2 (1+\zeta) \frac{\partial^2 e_{\rm xc}}{\partial n \partial \zeta}\nonumber\\
&&{}
+ \frac{1}{n}(1+\zeta)^2 \frac{\partial^2 e_{\rm xc}}{\partial \zeta^2}
\end{eqnarray}
\begin{eqnarray}
f^{\rm xc}_{\uparrow \uparrow,\downarrow \downarrow} =
f^{\rm xc}_{\downarrow \downarrow,\uparrow \uparrow}
&=& 2\: \frac{\partial e_{\rm xc}}{\partial n}
+ n\: \frac{\partial^2 e_{\rm xc}}{\partial n^2}
-2 \zeta \frac{\partial^2 e_{\rm xc}}{\partial n \partial \zeta} \nonumber\\
&&{}
- \frac{1}{n}(1-\zeta^2) \frac{\partial^2 e_{\rm xc}}{\partial \zeta^2}
\end{eqnarray}
\begin{equation}
f^{\rm xc}_{\ua\da,\ua\da} =
f^{\rm xc}_{\da\ua,\da\ua}=
\frac{2}{n\zeta} \: \frac{\partial e_{\rm xc}}{\partial \zeta} \:, \label{f4}
\end{equation}
where $e_{\rm xc}(n,\zeta)$ is the xc energy per particle of the spin-polarized HEG \cite{Perdew1992,Attaccalite2002}.

\section{Spin-flip response functions of the HEG: analytic results}\label{App:B}

In this Appendix we present the results for the real and imaginary parts of the 3D and 2D spin-flip response functions of the HEG.
The derivations are similar to those of the (spin-conserving) Lindhard functions of Ref. \cite{GiulianiVignale}, and we won't reproduce
the technical details here.

\subsubsection{3D case}

In 3D, the real parts of $\chi_{\ud,\ud}$ and $\chi_{\du,\du}$ are given by
\begin{widetext}
\begin{equation}
        \Re{\chi_{\uparrow\downarrow,\uparrow\downarrow}} = \frac{1}{2\pi^2}\biggl\{ \frac{k_{F\downarrow}^2}{4q} \biggl(1-(\nu_{-\downarrow}^{-\Delta})^2 \biggr)\ln{\left|\frac{\nu_{-\downarrow}^{-\Delta} + 1}{\nu_{-\downarrow}^{-\Delta} -1}\right|} + \frac{k_{F\downarrow}^2}{2q}\nu_{-\downarrow}^{-\Delta} - \frac{k_{F\uparrow}^2}{4q}\biggl(1-(\nu_{+\uparrow}^{-\Delta})^2\biggr)\ln{\left|\frac{\nu_{+\uparrow}^{-\Delta}+1}{\nu_{+\uparrow}^{-\Delta}-1}\right|} - \frac{k_{F\uparrow}^2}{2q}\nu_{+\uparrow}^{-\Delta}\biggr\}
\end{equation}
\begin{equation}
        \Re{\chi_{\downarrow\uparrow,\downarrow\uparrow}} = \frac{1}{2\pi^2}\biggl\{- \frac{k_{F\downarrow}^2}{4q} \biggl(1-(\nu_{+\downarrow}^{+\Delta})^2 \biggr)\ln{\left|\frac{\nu_{+\downarrow}^{+\Delta} + 1}{\nu_{+\downarrow}^{+\Delta} -1}\right|} - \frac{k_{F\downarrow}^2}{2q}\nu_{+\downarrow}^{+\Delta} + \frac{k_{F\uparrow}^2}{4q}\biggl(1-(\nu_{-\uparrow}^{+\Delta})^2\biggr)\ln{\left|\frac{\nu_{-\uparrow}^{+\Delta}+1}{\nu_{-\uparrow}^{+\Delta}-1}\right|} + \frac{k_{F\uparrow}^2}{2q}\nu_{-\uparrow}^{+\Delta}\biggr\}
\end{equation}
\end{widetext}
where \begin{equation}
    \nu_{\pm\sigma}^{\pm\Delta} = \frac{\omega}{qk_{F\sigma}} \pm \frac{q}{2k_{F\sigma}} \pm \frac{\Delta}{qk_{F\sigma}} \:.
\end{equation}
Here, the upper $\pm$ corresponds to the $\Delta/qk_{F\sigma}$ term and the lower $\pm$ corresponds to the $q/2k_{F\sigma}$ term. $k_{F\uparrow}$ and $k_{F\downarrow}$ are the spin-resolved Fermi wavevectors, defined as $k_{F\sigma} = k_F(1+\sigma\zeta)^{1/\tt d}$, where the Fermi energy $\varepsilon_F = k_F^2/2$ and $\zeta$ is the spin polarization. For the calculation of the spin wave stiffness, it is helpful to use response functions to the lowest-order wavevector. That is, we let $q \rightarrow 0$ such that $q \ll \omega$. In 3D, these are:
\begin{equation}
    \begin{aligned}
        \Re{\chi_{\uparrow\downarrow,\uparrow\downarrow}} = &-\frac{n\zeta}{\omega - \Delta} + \frac{nq^2/2}{(\omega - \Delta)^2} - \frac{(k_{F\uparrow}^5 - k_{F\downarrow}^5)q^2}{30\pi^2(\omega-\Delta)^3}
    \end{aligned}
\end{equation}
\begin{equation}
    \begin{aligned}
        \Re{\chi_{\downarrow\uparrow,\downarrow\uparrow}} = &\frac{n\zeta}{\omega + \Delta} + \frac{nq^2/2}{(\omega + \Delta)^2} + \frac{(k_{F\uparrow}^5 - k_{F\downarrow}^5)q^2}{30\pi^2(\omega+
        \Delta)^3}\:.
    \end{aligned}
\end{equation}
To compute $\Im{\chi_{\uparrow\downarrow,\uparrow\downarrow}}$ and $\Im{\chi_{\downarrow\uparrow,\downarrow\uparrow}}$, we use the relation
\begin{equation}\label{pv}
    \lim_{\eta\to0}(z-i\eta)^{-1} = P\biggl(\frac{1}{z}\biggr) + i\pi\delta(z) \:.
\end{equation}
After integration, we obtain the result:
\begin{equation}
    \begin{aligned}
        \Im{\chi_{\uparrow\downarrow,\uparrow\downarrow}} = \frac{1}{4\pi}\biggl\{&\frac{-k_{F\uparrow}^2}{2q}\biggl(1-\bigl(\nu_{+\uparrow}^{-\Delta}\bigr)^2\biggr)\Theta\biggl(1-\bigl(\nu_{+\uparrow}^{-\Delta}\bigr)^2\biggr) \\
        &\frac{-k_{F\downarrow}^2}{2q}\biggl(1-\bigl(\nu_{-\downarrow}^{-\Delta}\bigr)^2\biggr)\Theta\biggl(1-\bigl(\nu_{-\downarrow}^{-\Delta}\bigr)^2\biggr)\biggr\}
    \end{aligned}
\end{equation}
\begin{equation}
    \begin{aligned}
        \Im{\chi_{\downarrow\uparrow,\downarrow\uparrow}} = \frac{1}{4\pi}\biggl\{&\frac{-k_{F\downarrow}^2}{2q}\biggl(1-\bigl(\nu_{+\downarrow}^{+\Delta}\bigr)^2\biggr)\Theta\biggl(1-\bigl(\nu_{+\downarrow}^{+\Delta}\bigr)^2\biggr) \\
        &\frac{-k_{F\uparrow}^2\href{}{}}{2q}\biggl(1-\bigl(\nu_{-\uparrow}^{+\Delta}\bigr)^2\biggr)\Theta\biggl(1-\bigl(\nu_{-\uparrow}^{+\Delta}\bigr)^2\biggr)\biggr\}
    \end{aligned}
\end{equation}

\subsubsection{2D case}
In 2D, the real parts of $\chi_{\ud,\ud}$ and $\chi_{\du,\du}$ are given by
\begin{equation}
    \begin{aligned}
            &\Re{\chi_{\uparrow\downarrow,\uparrow\downarrow}} = \\
            \frac{1}{2\pi}\biggl\{&-1 + \sgn{\bigl(\nu_{+\uparrow}^{-\Delta}\bigr)}\Theta\biggl(\bigl(\nu_{+\uparrow}^{-\Delta}\bigr)^2-1\biggr)\frac{k_{F\uparrow}}{q}\sqrt{\bigl(\nu_{+\uparrow}^{-\Delta}\bigr)^2 -1} \\
            &-\sgn{\bigl(\nu_{-\downarrow}^{-\Delta}\bigr)}\Theta\biggl(\bigl(\nu_{-\downarrow}^{-\Delta}\bigr)^2-1\biggr)\frac{k_{F\downarrow}}{q}\sqrt{\bigl(\nu_{-\downarrow}^{-\Delta}\bigr)^2 -1}\biggr\} \\
    \end{aligned}
\end{equation}
\begin{equation}
    \begin{aligned}
        &\Re{\chi_{\downarrow\uparrow,\downarrow\uparrow}}= \\
        \frac{1}{2\pi}\biggl\{&-1 + \sgn{\bigl(\nu_{+\downarrow}^{+\Delta}\bigr)}\Theta\biggl(\bigl(\nu_{+\downarrow}^{+\Delta}\bigr)^2-1\biggr)\frac{k_{F\downarrow}}{q}\sqrt{\bigl(\nu_{+\downarrow}^{+\Delta}\bigr)^2 -1}
        \\
        &-\sgn{\bigl(\nu_{-\uparrow}^{+\Delta}\bigr)}\Theta\biggl(\bigl(\nu_{-\uparrow}^{+\Delta}\bigr)^2-1\biggr)\frac{k_{F\uparrow}}{q}\sqrt{\bigl(\nu_{-\uparrow}^{+\Delta}\bigr)^2 -1}\biggr\}. \\
    \end{aligned}
\end{equation}
Expanding in the lowest-order wavevector gives the expressions:
\begin{equation}
    \Re{\chi_{\uparrow\downarrow,\uparrow\downarrow}} = \frac{-n\zeta}{\omega - \Delta} + \frac{nq^2/2}{(\omega - \Delta)^2} - \frac{\pi\zeta n^2q^2}{(\omega - \Delta)^3}
\end{equation}
\begin{equation}
    \Re{\chi_{\downarrow\uparrow,\downarrow\uparrow}} = \frac{n\zeta}{\omega + \Delta} + \frac{nq^2/2}{(\omega + \Delta)^2} + \frac{\pi\zeta n^2q^2}{(\omega+\Delta)^3} \:.
\end{equation}
For the imaginary part, we use the same relation (\ref{pv}) as in the 3D case, which gives the result:
\begin{equation}
    \begin{aligned}
        \Im{\chi_{\uparrow\downarrow,\uparrow\downarrow}} = \frac{1}{2\pi}\biggl\{ &\frac{k_{F\uparrow}}{q}\Theta\biggl(1-\bigl(\nu_{+\uparrow}^{-\Delta}\bigr)^2\biggr)\sqrt{1-\bigl(\nu_{+\uparrow}^{-\Delta}\bigr)^2}\\
         -&\frac{k_{F\downarrow}}{q}\Theta\biggl(1-\bigl(\nu_{-\downarrow}^{-\Delta}\bigr)^2\biggr)\biggr\}\sqrt{1-\bigl(\nu_{-\downarrow}^{-\Delta}\bigr)^2}\biggr\} \\
    \end{aligned}
\end{equation}
\begin{equation}
    \begin{aligned}
        \Im{\chi_{\downarrow\uparrow,\downarrow\uparrow}} = \frac{1}{2\pi}\biggl\{ &\frac{k_{F\downarrow}}{q}\Theta\biggl(1-\bigl(\nu_{+\downarrow}^{+\Delta}\bigr)^2\biggr)\sqrt{1-\bigl(\nu_{+\downarrow}^{+\Delta}\bigr)^2}\\
         -&\frac{k_{F\uparrow}}{q}\Theta\biggl(1-\bigl(\nu_{-\uparrow}^{+\Delta}\bigr)^2\biggr)\biggr\}\sqrt{1-\bigl(\nu_{-\uparrow}^{+\Delta}\bigr)^2}\biggr\}
    \end{aligned}
\end{equation}

\section{Source-free linear response formalism}\label{App:C}

In a spin-polarized HEG, the xc magnetic field
\begin{equation}
\bfB_{\rm xc}^{\rm HEG} = \frac{\bfm}{n\zeta} \frac{\partial e_{\rm xc}}{\partial \zeta}
\end{equation}
is uniform and hence source-free by default. To include this case in the definition of the functional $\tilde E_{\rm xc}$ (see Sec. \ref{Sec:2A}) requires
some care, as discussed in Ref. \cite{Sharma2018}, and the Helmholtz construction (\ref{B_SF}) is not really meaningful.
In the following, we will consider a non-uniform system within the LSDA, derive the source-free construction,
and only in the end take the uniform limit. As we will see, the SF construction will give rise to additional terms that do not
vanish in this limit.

In LSDA, the source-free magnetic field (ignoring here the empirical scaling factor $s$ for simplicity)
is given by the transverse component of the $\bfB_{\rm xc}$ vector field:
\begin{equation}
\bfB_{\rm xc,SF}^{\rm LSDA}(\bfr) = \frac{1}{4\pi} \nabla \times \int d\bfr' \:\frac{\nabla' \times \bfB_{\rm xc}^{\rm LSDA}(\bfr')}{|\bfr - \bfr'|}
\end{equation}
or, explicitly for the LSDA,
\begin{equation}
\bfB_{\rm xc,SF}^{\rm LSDA}(\bfr) = \frac{1}{4\pi} \nabla \times \int d\bfr' \frac{\nabla' \times \frac{\bfm(\bfr') e_{\rm xc} '(\bfr')}{n(\bfr') \zeta(\bfr')}}
{|\bfr - \bfr'|} \:,
\end{equation}
where we use the short-hand notation
\begin{equation}\label{exc_prime}
\exc'(\bfr) = \left.\frac{\partial e_{\rm xc}(n,\zeta)}{\partial \zeta} \right|_{\genfrac{}{}{0pt}{}{n = n(\bfr)}{\zeta = \zeta(\bfr)}}
\end{equation}
and we introduce the abbreviation
\begin{equation}
g_j(\bfr) =  \frac{m_j(\bfr) \exc'(\bfr)}{n(\bfr) \zeta(\bfr)} \:, \qquad j=1,2,3.
\end{equation}
Now let us calculate the xc kernel $h_{jk}^{\rm xc}$. We first consider the components with $j,k = 1,2,3$, where
\begin{equation}
h_{jk}^{\rm xc}(\bfr,\bfr') = \frac{\delta B_{{\rm xc},j}(\bfr)}{\delta m_k(\bfr')}.
\end{equation}
We obtain

\begin{widetext}

\begin{equation}
h_{jk}^{\rm xc}(\bfr,\bfr') =
\frac{\epsilon_{jmn}\epsilon_{npq} }{4\pi}
\nabla_m  \int d\bfr'' \frac{\nabla''_p  \delta (\bfr'' - \bfr')}{|\bfr - \bfr''|}
\left[ \frac{\delta_{qk}e_{\rm xc}'(\bfr')}{n(\bfr') \zeta(\bfr')}
- \frac{m_q(\bfr')m_k(\bfr')}{n^3(\bfr')\zeta^3(\bfr')}(e_{\rm xc}'(\bfr') - \zeta(\bfr')e_{\rm xc}''(\bfr'))\right], \nonumber
\end{equation}
where $\epsilon_{jkl}$ is the Levi-Civita symbol, and $e_{\rm xc}''$ is the second derivative of $\exc$ with respect to $\zeta$, similarly to $\exc'$ defined
in Eq. (\ref{exc_prime}). In the following, it will be convenient to express the integral over the delta function
via Fourier transformation:
\begin{equation}
\nabla_m  \int d\bfr'' \frac{\nabla''_p  \delta (\bfr'' - \bfr')}{|\bfr - \bfr''|}
=
-4\pi \int \frac{d\bfq}{(2\pi)^3} \frac{q_m q_p}{q^2} e^{i\bfq(\bfr - \bfr')} \:.
\end{equation}
After some manipulation one then arrives at
\begin{eqnarray}\label{hxcij}
h_{jk}^{\rm xc}(\bfr,\bfr')
&=&
\left[ \frac{\delta_{jk}e_{\rm xc}'(\bfr)}{n(\bfr) \zeta(\bfr)}
- \frac{m_j(\bfr)m_k(\bfr)}{n^3(\bfr)\zeta^3(\bfr)}(e_{\rm xc}'(\bfr) - \zeta(\bfr)e_{\rm xc}''(\bfr))\right]
\delta(\bfr - \bfr') \nonumber\\
&-& \sum_n
\left[ \frac{\delta_{nk}e_{\rm xc}'(\bfr')}{n(\bfr') \zeta(\bfr')}
- \frac{m_n(\bfr')m_k(\bfr')}{n^3(\bfr')\zeta^3(\bfr')}(e_{\rm xc}'(\bfr') - \zeta(\bfr')e_{\rm xc}''(\bfr'))\right]
\int \frac{d^3q}{(2\pi)^3} \frac{q_n q_j}{q^2} e^{i\bfq(\bfr - \bfr')}.
\end{eqnarray}
The first part is the collinear adiabatic LSDA, the second part is a new source-free correction.

We now consider the case of a spin-polarized HEG where $m_1=m_2=0$, and $m_3=\zeta n$.
From Eq. (\ref{hxcij}) we then obtain a $\bfq$-dependent kernel:
\begin{equation}
h_{jk}^{\rm xc}(\bfq)
=
\delta_{jk}\left[ \frac{e_{\rm xc}'}{n \zeta}
- \frac{\delta_{j3}}{n\zeta}(e_{\rm xc}' - \zeta e_{\rm xc}'')\right]
-
\left[ \frac{e_{\rm xc}'}{n \zeta}
- \frac{\delta_{k3}}{n\zeta}(e_{\rm xc}' -\zeta e_{\rm xc}'')\right] \frac{q_k q_j}{q^2}
\end{equation}
We also need the derivatives with respect to the scalar density. $h^{\rm xc}_{00}$ and $h^{\rm xc}_{0k}$ are unchanged with respect to the LSDA, but
we get new results for $h_{j0}^{\rm xc}$, $j=1,2,3$.
Specifically, we find
\begin{equation}
h_{j0}^{\rm xc}(\bfr,\bfr')
=
-\left[ \frac{m_q(\bfr')}{n(\bfr')\zeta(\bfr')} \frac{\partial^2 e_{\rm xc}}{\partial n\partial \zeta}
- \frac{ m_q(\bfr')}{n^2(\bfr')} \frac{ \partial^2 e_{\rm xc}}{\partial \zeta^2} \right]
\epsilon_{jmn}\epsilon_{npq} \int \frac{d^3q}{(2\pi)^3} \frac{q_m q_p}{q^2} e^{i\bfq(\bfr - \bfr')}
\end{equation}
and after some manipulation we end up with
\begin{eqnarray}
h_{j0}^{\rm xc}(\bfr,\bfr')
&=&
\left[ \frac{m_j(\bfr)}{n(\bfr)\zeta(\bfr)} \frac{\partial^2 e_{\rm xc}}{\partial n\partial \zeta}
- \frac{ m_j(\bfr)}{n^2(\bfr)} \frac{ \partial^2 e_{\rm xc}}{\partial \zeta^2} \right] \delta(\bfr - \bfr')\\
&-&
\sum_n \left[ \frac{m_n(\bfr')}{n(\bfr')\zeta(\bfr')} \frac{\partial^2 e_{\rm xc}}{\partial n\partial \zeta}
- \frac{ m_n(\bfr')}{n^2(\bfr')}\frac{ \partial^2 e_{\rm xc}}{\partial \zeta^2} \right] \int \frac{d^3q}{(2\pi)^3}\frac{q_n q_j}{q^2}e^{i\bfq(\bfr - \bfr')} \:.
\end{eqnarray}
Again, first part is the collinear LSDA. For the HEG with $m_1=m_2=0$, and $m_3=\zeta n$, one then obtains the $\bfq$-dependent kernel
\begin{equation}
h_{j0}^{\rm xc}(\bfq)
=
\left( \delta_{j3} - \frac{q_3 q_j}{q^2}\right)\left[  \frac{\partial^2 e_{\rm xc}}{\partial n\partial \zeta}
- \frac{ \zeta}{n} \frac{ \partial^2 e_{\rm xc}}{\partial \zeta^2} \right].
\end{equation}
Let us now write the full xc kernel of the HEG, including source-free correction, in matrix form:
\begin{equation} \label{hxc_matrix}
\underline{\underline h}^{\rm xc}(\bfq) =  \left( \begin{array}{cccc}
h^{\rm xc}_{00}& 0  &  0 &\frac{\partial^2 e_{\rm xc}}{\partial n\partial \zeta}- \frac{ \zeta e_{\rm xc}''}{n}  \\[2mm]
-\left[\frac{\partial^2 e_{\rm xc}}{\partial n\partial \zeta}- \frac{ \zeta e_{\rm xc}''}{n}  \right]\frac{q_1 q_3}{q^2} &\frac{e_{\rm xc}'}{n\zeta} (1-\frac{q_1^2}{q^2}) &  -\frac{e_{\rm xc}'}{n\zeta} \frac{q_1 q_2}{q^2}   &  -\frac{e_{\rm xc}''}{n} \frac{q_1 q_3}{q^2} \\[2mm]
-\left[\frac{\partial^2 e_{\rm xc}}{\partial n\partial \zeta}- \frac{ \zeta e_{\rm xc}''}{n}  \right]\frac{q_2 q_3}{q^2} &-\frac{e_{\rm xc}'}{n\zeta} \frac{q_1 q_2}{q^2}    & \frac{e_{\rm xc}'}{n\zeta}(1 - \frac{q_2^2}{q^2}) &   -\frac{e_{\rm xc}''}{n} \frac{q_2 q_3}{q^2}  \\[2mm]
\left[\frac{\partial^2 e_{\rm xc}}{\partial n\partial \zeta}- \frac{ \zeta e_{\rm xc}''}{n}  \right](1 - \frac{q_3^2}{q^2})
&-\frac{e_{\rm xc}'}{n\zeta} \frac{q_1 q_3}{q^2}    &  -\frac{e_{\rm xc}'}{n\zeta} \frac{q_2 q_3}{q^2} &   \frac{e_{\rm xc}''}{n}(1 - \frac{q_3^2}{q^2})
\end{array}\right).
\end{equation}
This matrix is obviously not symmetric. Since the xc kernel is defined as a second functional derivative, it should have the exact property
\begin{equation}
h^{\rm xc}_{jk}(\bfr,\bfr') = h^{\rm xc}_{kj}(\bfr',\bfr)
\end{equation}
and so we would expect here that $h^{\rm xc}_{jk}(\bfq) = h^{\rm xc}_{kj}(-\bfq)$. Clearly, expression (\ref{hxc_matrix}) does not behave like that.
The reason is obvious: the source-free xc magnetic field is not variational (it is not a functional derivative).
A quick and easy fix would be to symmetrize the matrix by hand, setting $\underline{\underline h}^{\rm xc}_{\rm symm} = \frac{1}{2}\underline{\underline h}^{\rm xc}
+ \frac{1}{2}(\underline{\underline h}^{\rm xc})^T$.

However, we are here considering a special case where the symmetry violation does not occur. Namely,
we are considering spin waves that propagate perpendicular to the applied
magnetic field (which sets the quantization axis). Hence, assuming that the magnetic field points along $z$ (or 3), we can set
$q_3=0$, and only consider a wavevector $\bfq_{\perp}$ that is in the plane perpendicular to the external magnetic field. This simplifies
things enormously, and we get
\begin{equation}
\underline{\underline h}^{\rm xc}(\bfq_\perp) =  \left( \begin{array}{cccc}
h^{\rm xc}_{00}& 0 & 0 &  h^{\rm xc}_{03}
\\[2mm]
0 &(1-\frac{q_1^2}{q^2})h^{\rm xc}_{11}  &  -\frac{q_1 q_2}{q^2} h^{\rm xc}_{11}  & 0 \\[2mm]
0 &- \frac{q_1 q_2}{q^2}h^{\rm xc}_{11}     & (1 - \frac{q_2^2}{q^2})h^{\rm xc}_{11}  & 0 \\[2mm]
h^{\rm xc}_{03} & 0   &  0    &  h^{\rm xc}_{33}
\end{array}\right).
\end{equation}
Transforming this using Eq. (\ref{kerneltrafo}) finally leads to
\begin{equation}\label{fxc_sf_mat}
\underline{\underline f}^{\rm xc}(\bfq_\perp)
=
\left( \begin{array}{cccc}
h^{\rm xc}_{00} + 2h^{\rm xc}_{03} + h^{\rm xc}_{33}& 0 & 0 &  h^{\rm xc}_{00} - h^{\rm xc}_{33}
\\[2mm]
0 & h^{\rm xc}_{11}  &  \frac{(q_2 + i q_1)^2}{q^2} h^{\rm xc}_{11}  & 0 \\[2mm]
0 & \frac{(q_2 - iq_1)^2}{q^2} h^{\rm xc}_{11}   & h^{\rm xc}_{11}  & 0 \\[2mm]
h^{\rm xc}_{03} - h^{\rm xc}_{33} & 0   &  0    &  h^{\rm xc}_{00} - 2 h^{\rm xc}_{03} + h^{\rm xc}_{33}
\end{array}\right).
\end{equation}

\end{widetext}
\bibliography{SF_refs}
\end{document}